\documentclass[twocolumn,prl,aps,floats]{revtex4}

\usepackage{epsf,graphicx,xcolor}
\usepackage{supertabular}
\usepackage{epsfig,amsmath}

\newcommand{\ci}{\hspace{-0.6pt}\circ\hspace{-0.6pt}}
\newcommand{\ov}{\overline}

\begin{document}

\frenchspacing
\title{Exact dynamics of the critical Kauffman model with connectivity one}
\author{T.~M.~A.~Fink}
\affiliation{London Institute for Mathematical Sciences, Royal Institution, 21 Albemarle St, London W1S 4BS, UK}

\date{\today}
\begin{abstract}
\noindent
The critical Kauffman model with connectivity one is the simplest class of critical Boolean networks.
Nevertheless, it exhibits intricate behavior at the boundary of order and chaos.
We introduce a formalism for expressing the dynamics of multiple loops as a product of the dynamics of individual loops.
Using it, we prove that the number of attractors scales as $2^m$, where $m$ is the number of nodes in loops---as fast as possible, and much faster than previously believed.
\end{abstract}

\maketitle

\noindent
Exactly solvable models play a special role in physics.
They allow us to understand why a model behaves the way that it does.
They also suggest lines of attack for more realistic models which cannot be solved exactly \cite{Flyvbjerg88}.
For some exact models, new approaches to the problem continue to reveal additional structure and insights.
\\ \indent
One such model is the critical Kauffman network with connectivity one. 
It is a Boolean network in which the state of each node is a logical function of just one other node.
The network is thus composed of loops and trees branching off of the loops, as shown in Fig. 1.
Because the nodes in the trees are slaves to the loops, they do not contribute to the number or length of attractors, which are set solely by the $m$ nodes in the loops.
\\ \indent
The model was first studied comprehensively in 1988 by Flyvbjerg and Kjaer in a 24 page paper \cite{Flyvbjerg88}.
Then, 17 years later, Drossel, Mihaljev and Greil \cite{Drossel05} obtained a more complete understanding of the critical behavior by generating networks through a growth process.
Perhaps it is fitting that, 17 years after that, we take a more direct approach to reveal qualitatively new results. 
\\ \indent
In a general critical Kauffman model \cite{Bilke02, Socolar03, Troein03, Peixoto10},
 a perturbation to one node propagates to, on average, one other node \cite{Munoz18}.
Since in our model every node has one input, a node can apply one of four Boolean functions: on, off, copy and invert.
However, the critical version of the problem, which we study here, requires that all Boolean functions in the loops be copy or invert---just one on or off freezes the loop, rendering it irrelevant.
\\ \indent
Our goal is to calculate the number of attractors. 
We start by writing down the exact dynamics or a single loop, then multiple loops of the same size.
(Throughout, we use the word dynamics to refer to the number and size of attractors.)
We then introduce a formalism for expressing the dynamics of loops of different sizes as a product of the dynamics of loops of the same size.
Using this, we prove that the number of attractors scales as $2^m$, where $m$ is the number of nodes in loops,
considerably faster than the best known bounds of $2^{0.5m}$ and $2^{0.47m}$ derived in \cite{Flyvbjerg88} and \cite{Drossel05}.
This is the first proof that the number of attractors grows as fast as possible with $m$.
In terms of the network size $N$, it scales at least as fast as $2^{1.25 \sqrt{N}}$,
compared to $2^{0.63 \sqrt{N}}$ and $2^{0.59 \sqrt{N}}$ \cite{Flyvbjerg88, Drossel05}.
We also prove that the mean attractor length is at most $2^{\sqrt{2m} \log_2\!\sqrt{2m}}$.
\begin{figure}[b!]
\noindent
      \epsfxsize=1 \columnwidth
      \epsfbox{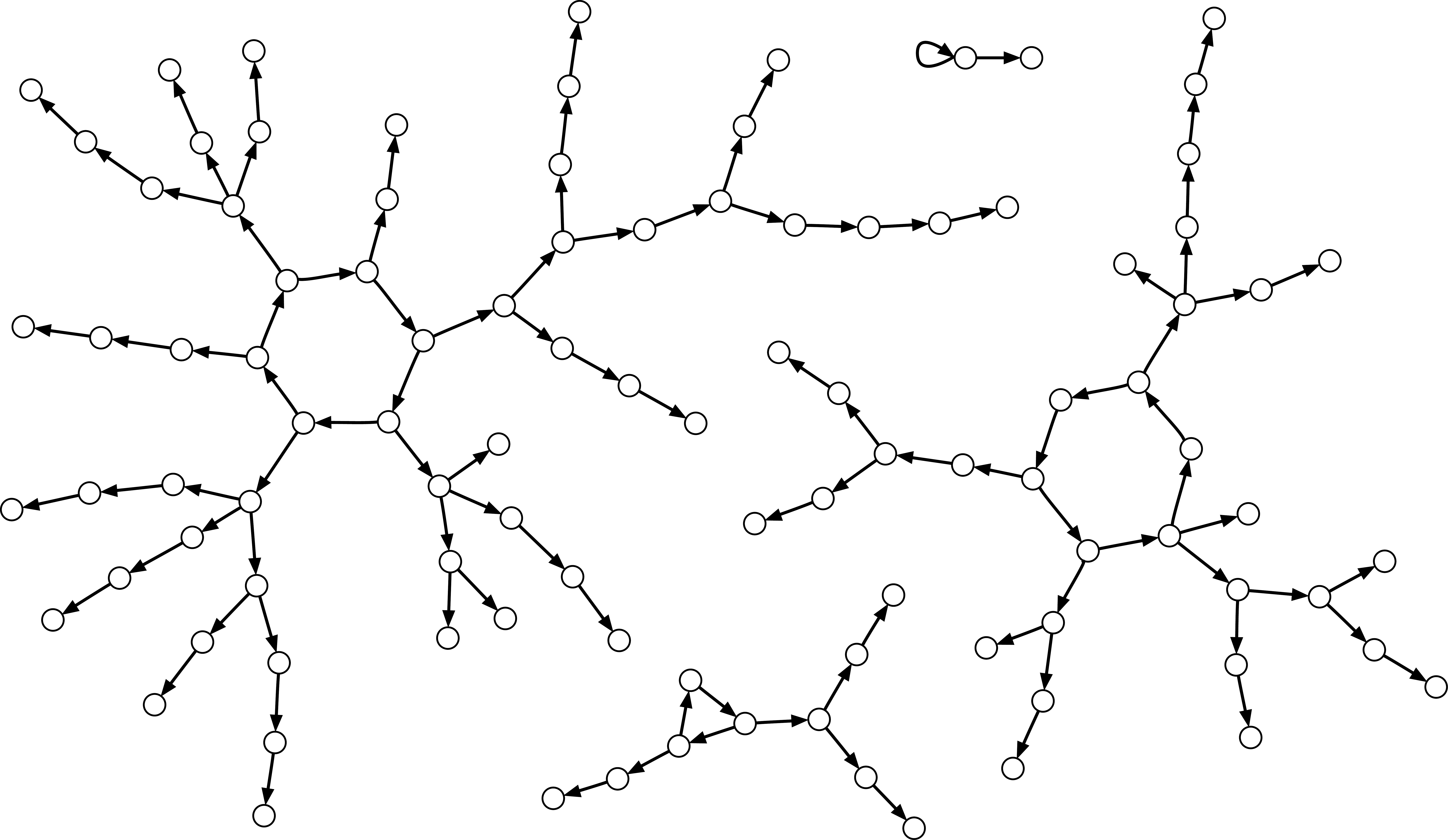}
\begin{small}
\caption{
{\bf Kauffman network with connectivity one.} 
This typical network of $N=100$ nodes has one 1-loop, one 3-loop and two 6-loops, for a total of $m=16$ nodes in loops.
}
\end{small}
\end{figure}
\begin{table*}[t!]
\noindent
\begin{tabular*}{\textwidth}{@{\extracolsep{\fill}}llll}
$d(1) = 2 \ci 1$ 							& $d(\ov{1}) = 1 \ci 2$ 			& $d(1^2) = 4 \ci 1$ 							& $d(\ov{1}^2) = 2 \ci 2$ 				\\ 
$d(2) = 2 \ci 1 + 1 \ci 2$ 					& $d(\ov{2}) = 1 \ci 4$			& $d(2^2) = 4 \ci 1 + 6 \ci 2$ 					& $d(\ov{2}^2) = 4 \ci 4$				\\
$d(3) = 2 \ci 1 + 2 \ci 3$ 					& $d(\ov{3}) = 1 \ci 2 + 1 \ci 6$		& $d(3^2) = 4 \ci 1 + 20 \ci 3$ 					& $d(\ov{3}^2) = 2 \ci 2 + 10 \ci 6$		\\ 	
$d(4) = 2 \ci 1 + 1 \ci 2 + 3 \ci 4$ 			& $d(\ov{4}) = 2 \ci 8$			& $d(4^2) = 4 \ci 1 + 6 \ci 2 + 60 \ci 4$ 			& $d(\ov{4}^2) = 32 \ci 8$				\\
$d(5) = 2 \ci 1 + 6 \ci 5$ 					& $d(\ov{5}) = 1 \ci 2 + 3 \ci 10$		& $d(5^2) = 4 \ci 1 + 204 \ci 5$ 					& $d(\ov{5}^2) = 2 \ci 2 + 102 \ci 10$		\\
$d(6) = 2 \ci 1 + 1 \ci 2 + 2 \ci 3 + 9 \ci 6$ 		& $d(\ov{6}) = 1 \ci 4 + 5 \ci 12$		& $d(6^2) = 4 \ci 1 + 6 \ci 2 + 20 \ci 3 + 670 \ci 6$ 	& $d(\ov{6}^2) = 4 \ci 4 + 340 \ci 12$		\\
$d(7) = 2 \ci 1 + 18 \ci 7$ 					& $d(\ov{7}) = 1 \ci 2 + 9 \ci 14$		& $d(7^2) = 4 \ci 1 + 2340 \ci 7$ 				& $d(\ov{7}^2) = 2 \ci 2 + 1170 \ci 14$	\\
$d(8) = 2 \ci 1 + 1 \ci 2 + 3 \ci 4 + 30 \ci 8$ 	& $d(\ov{8}) = 16 \ci 16$			& $d(8^2) = 4 \ci 1 + 6 \ci 2 + 60 \ci 4 + 8160 \ci 8$ 	& $d(\ov{8}^2) = 4096 \ci 16$		
\end{tabular*}
\caption{
{\bf Dynamics for loops of the same size.} 
The dynamics $d(l)$ of an even loop of length $l$ has $a_k$ cycles of length $k$ if $k$ divides $l$, where, e.g.,
$d(2)$ reads as 2 cycles of length 1 and 1 cycle of length 2.
The dynamics of an odd loop has $b_k$ cycles of length $k$ if $k$ divides $2l$ but not $l$. 
For $n$ loops of the same size, the same applies but with $a_k$ and $b_k$ replaced by $a_{k,n}$ and $b_{k,n}$.
}
\end{table*}
\\ \noindent
{\sf\textbf{\textcolor{purple}{Single loop}}} \\
For a loop of length $l$, there are $2^l$ ways of assigning copy and invert 
to the $l$ nodes, but these lead to just two behaviors \cite{Kaufman05}.
If the number of inverts is even, 
the dynamics is identical to all of them being copy; this is called an even loop.
If the number of inverts is odd, the dynamics is identical to all of them being copy apart from one invert; 
this is called an odd loop.
We can express the dynamics of even and odd loops in terms of the sequences
	\begin{equation*}
		a_k = {1\over k} \sum_{j\vert k} \mu(j) 2^{k/j}, 
		\qquad
		b_k = {1\over 2 k} \!\! \sum_{{\rm odd} \, j\vert k} \!\! \mu(j) 2^{k/j}
		\label{samesizeloopab}
	\end{equation*}
(OEIS A1037, A48 \cite{Sloane}),
where $\mu$ is the M\"{o}bius function.
The $a_k$ are the number of necklaces of $k$ beads in two colors with primitive period $k$.
The $b_k$ are the number of such necklaces when the two colors can be interchanged.
For odd prime $p$, $a_p = \textstyle (2^{p} - 2)/p$ and $b_{p} = (2^{p} - 2)/(2p)$.
\\ \indent
An even $l$-loop has cycles of length $k$ if and only if $k$ divides $l$; there are $a_k$ of them.
An odd $l$-loop has cycles of length $2 k$ if and only if $k$ divides $2 l$ but not $l$; there are $b_k$ of them.
We use the shorthand $\{ l \}$ and $\{ \overline{l} \}$ for even and odd loops of length $l$.
Let $i \ci j$ denote $i$ copies of a $j$-cycle.
Then we can write the dynamics as
\begin{equation*}
d(l) = \sum_{k|l} a_k \ci k, \qquad 
d(\overline{l}) =  \!\!\! \sum_{k\vert 2 l, \, k \not \hspace{2pt} \vert \, l } \!\!\! b_k \ci k,
\label{E}
\end{equation*}
where we take ``+'' to mean ``and'',
and we drop the braces around $l$ and $\overline{l}$ inside functions.
Examples are given in Table I.
Since all $2^l$ states of the loop belong to cycles, the sums of $a_k k$ and $b_k k$ are both $2^l$.
\\ \noindent {\sf\textbf{\textcolor{purple}{Multiple loops of the same size}}} \\
We call $n$ loops of length $l$ a cluster of $l$-loops.
Let the cluster have $p \geq 0$ even loops and $q = n-p \geq 0$ odd loops,
which we denote by $\{ l^p, \overline{l}^q \}$.
If all the loops in the cluster are even, we call it an even cluster.
If one or more loops is odd, then the dynamics is the same as if all loops were odd, and we call it an odd cluster.
Thus for $q \geq 1$, 
	\begin{equation}
		d(l^p\,\overline{l}^{q}) = d(\overline{l}^{\,p+q}). 
		\label{cluster_identity}
	\end{equation}

This can be seen as follows.
Consider the $n$ loops as concentric circles, and let $u^i_j$ be the value of the $j$th node in loop $i$.
Let $\alpha_j$ be the values of a radial cut through the circles:
$\alpha_j = u^1_j, \dots, u^n_j$.
Assume all $n$ loops are even, with no inverts.
Then $\alpha_j$ is just copied around the loops.
Now assume all loops are even except the first, with a single invert.
A first pass around the loops maps all the $\alpha_j$ to $\overline{u}^1_j, \dots, u^n_j$ which on a second pass is mapped back to $\alpha_j$
(here $\overline{u}$ means not $u$).
The same is true for any combination of the loops in which one or more are odd. 

Since $\alpha_j$ is in one of $2^n$ states, we can think of the cluster of loops as a single loop in which each node can take $2^n$ states.
We can write the dynamics of even and odd clusters as
\begin{equation*}
d(l^t) = \sum_{k|l} a_{k,n} \ci k, \qquad 
d(\overline{l}^{\,n}) =  \!\!\! \sum_{k\vert 2 l, \, k \not \hspace{2pt} \vert \, l } \!\!\! b_{k,n} \ci k,
\label{E}
\end{equation*}
where 
\begin{equation*}
    a_{k,n} = {1\over k} \sum_{j\vert k} \mu(j) 2^{n k/j}, \qquad 
    b_{k,n} = {1\over 2 k} \!\! \sum_{{\rm odd} \, j\vert k} \!\! \mu(j) 2^{n k/j}.
    \label{samesizeloopabl}
\end{equation*}
Examples are given in Table I.
The $a_{k,n}$ are the number of necklaces of $k$ beads in $2^n$ colors with primitive period $k$.
The $b_{k,n}$ are the number of such necklaces when each color can be interchanged with a unique other color.
\\ \indent
Let $c$ be the number of cycles in a cluster:
$c(l^{n}) = \sum_{k\vert l} a_{k,n}$ and $c(\overline{l}^{\, n}) = \sum_{k\vert 2 l, \, k \not \hspace{2pt} \vert \, l } b_{k,n}$.
We can re-express these using the identity 
	$\phi(k) = \sum_{j\vert k} j \, \mu(k/j)$,
where $\phi(k)$ is the Euler totient function.
We find 	
\begin{table*}[t!]
\noindent
\begin{tabular*}{\textwidth}{@{\extracolsep{\fill}}llll}
$d(1,2) 	= 4 \ci 1 + 2 \ci 2$				& $d(1,\ov{2}) 	= 2 \ci 4$			& $d(\ov{1},2) 	= 4 \ci 2$						& $d(\ov{1},\overline{2}) 	= 2 \ci 4$ \\
$d(1,3) 	= 4 \ci 1 + 4 \ci 3$				& $d(1,\ov{3}) 	= 2 \ci 2 + 2 \ci 6$	& $d(\ov{1},3) 	= 2 \ci 2 + 2 \ci 6$				& $d(\ov{1},\ov{3}) 	= 2 \ci 2 + 2 \ci 6$ \\
$d(2,3)  	= 4 \ci 1 + 2 \ci 2 + 4 \ci 3 + 2 \ci 6$	& $d(2,\ov{3}) 	= 4 \ci 2 + 4 \ci 6$	& $d(\ov{2},3) 	= 2 \ci 4 + 2 \ci 12$				& $d(\ov{2},\ov{3}) 	= 2 \ci 4 + 2 \ci 12$ \\
$d(1,4) 	= 4 \ci 1 + 2 \ci 2 + 6 \ci 4	$		& $d(1,\ov{4}) 	= 4 \ci 8$			& $d(\ov{1},4) 	= 4 \ci 2 + 6 \ci 4$				& $d(\ov{1},\ov{4}) 	= 4 \ci 8$ \\
$d(2,4) 	= 4 \ci 1 + 6 \ci 2 + 12 \ci 4$		& $d(2,\ov{4}) 	= 8 \ci 8$			& $d(\ov{2},4) 	= 16 \ci 4$						& $d(\ov{2},\ov{4}) 	= 8 \ci 8$ \\
$d(3,4) 	= 4 \ci 1 + 2 \ci 2 + 4 \ci 3$ 		& $d(3,\ov{4}) 	= 4 \ci 8 + 4 \ci 24$	& $d(\ov{3},4) 	= 4 \ci 2 + 6 \ci 4$				& $d(\ov{3},\ov{4}) 	= 4 \ci 8 + 4 \ci 24$ \\
	\hspace{24.4pt} $+ \,\, 6 \ci 4 + 2 \ci 6 + 6 \ci 12$ 	&					& \hspace{24.4pt} $+ \,\, 4 \ci 6 + 6 \ci 12$ 		
\end{tabular*}
\caption{
{\bf Dynamics for loops of different sizes.} 
These can be deduced from the dynamics of individual loops by the product formula in eq. (\ref{product}).
Here, e.g., $d(2,\ov{3})$ indicates the dynamics of an even 2-loop and an odd 3-loop.
}
\end{table*}
  \begin{equation*}
    c(l^n) = {1\over l} \sum_{k\vert l} \phi(k) 2^{n l / k}, \qquad 
    c(\overline{l}^{\, n}) = {1\over 2 l} \!\! \sum_{{\rm odd} \, k\vert l} \!\! \phi(k) 2^{n l / k}
    \label{samesizeloopAB}
  \end{equation*}
(OEIS A31, A16 \cite{Sloane}).
Taking just the $k=1$ term gives the following good bounds, which we will use later:
  \begin{equation}
    c(l^{n}) >  2^{n l}/l, \qquad 
    c(\overline{l}^{\, n}) > 2^{n l}/(2 l).
    \label{BoundsClusters}
  \end{equation}
 {\sf\textbf{\textcolor{purple}{Multiple loops of different sizes}}} \\
Loops of different sizes can give rise to more complicated behavior, where the cycle lengths of the network 
are the least common multiples of the cycle lengths of individual loops.
The dynamics of multiple loops can be deduced from that of individual loops by defining a product between dynamics:
\begin{eqnarray}
	(g_1 \ci x_1 + g_2 \ci x_2 + \ldots) (h_1 \ci y_1 + h_2 \ci y_2 + \ldots) \nonumber \\
  = \sum_{i,j} g_i h_j \, \gcd(x_i, y_j) \ci {\rm lcm}(x_i, y_j).
   \label{product}
\end{eqnarray}
\indent
Consider a collection of loops $\Theta$ in which there are $s$ loop sizes and therefore $s$ clusters.
For a given cluster, there are $n_i$ loops of size $l_i$, of which $p_i$ are even and $q_i = n_i - p_i$ are odd. 
From eq. (\ref{product}), we see that the number of cycles $c$ is super-multiplicative.
In particular,
\begin{equation}
c(l_1^{p_1},\overline{l}_1^{\, q_1} \ldots l_s^{p_s},\overline{l}_s^{\, q_s}) \geq 
\prod_{i=1}^s	c\big(l_i^{\,p_i}\big)
\prod_{i=1}^s 	c\big(\overline{l}_i^{\,q_i}\big).
\label{SuperAddPos}
\end{equation}
Since for odd clusters all cycle lengths are even,
\begin{equation}
c(\overline{l}_1^{\, q_1}\ldots\overline{l}_s^{\, q_s}) \geq 
2^{s-1} \prod_{i=1}^s c\big(\overline{l}_i^{\,q_i}\big).
\label{SuperAddNeg}
\end{equation}
 {\sf\textbf{\textcolor{purple}{Minimum number of cycles}}} \\
Equipped with the above results, we can now calculate the minimum number of cycles for $m$ nodes in loops.
We divide the $s$ clusters into two categories: 
those in which one or more of the loops is odd---of which there are some number $r$---and those in which they are all even:
	\begin{equation*}
		c(\Theta) = c \big(l_1^{p_1}, \overline{l}_1^{\,q_1}, \ldots, l_q^{p_r}, \overline{l}_r^{\,q_r}, 	l_{r+1}^{n_{r+1}}, \ldots, l_s^{n_s}\big). 
	\end{equation*}
Assume at least one of the clusters in $\Theta$ is odd (we will deal with the alternative case below).
By eq. (\ref{cluster_identity}),
	\begin{equation*}
		c(\Theta) = c \big(\overline{l}_1^{\, n_1}, \ldots,  \overline{l}_r^{\, n_r}, l_{r+1}^{n_{r+1}}, \ldots, l_s^{n_s}\big).
	\end{equation*}
Applying the inequalities in eqs. (\ref{SuperAddPos}) and (\ref{SuperAddNeg}),
	\begin{equation*}
		c(\Theta) 	\geq 	2^{r-1} \prod_{i=1}^r c\!\left(\overline{l}_i^{\,n_i}\right) \prod_{i=r+1}^s	c\!\left(l_i^{\,n_i}\right).
	\end{equation*}
Using the bounds in eq. (\ref{BoundsClusters}), and since $\sum_{i = 1}^s n_i l_i = m$,
	\begin{equation}
		c(\Theta) > 2^{m-1} \prod_{i=1}^s \frac{1}{l_i}.
		\label{OverEll}
	\end{equation}
If no loop in $\Theta$ is odd, the bound is twice this.
\\ \indent
To minimize the right side of eq. (\ref{OverEll}), we want to maximize the product of the $l_i$, which occurs when the $n_i$ are all 1, that is, $\sum_{i = 1}^s l_i = m$.
We want the distinct $l_i$ as small as possible but greater than 1.
For $m = {(s+1)(s+2) \over 2} - 1$, the optimal choice of the $l_i$ is $2, 3, \ldots, s+1$.
As $m$ increases, this sequence progresses by incrementing one element at a time, from right to left.
The process restarts after the leftmost element is incremented.
For example, for $s=3$ and $m=9$, the progression is: 
2,3,4; 2,3,5; 2,4,5; 3,4,5; 3,4,6; and so on.
When $m$ reaches ${(s+2)(s+3) \over 2} - 1$, the number of $l_i$ increases from $s$ to $s+1$.
Thus $\prod l_i$ is at most $\prod_{i=2}^{s+1} i$ 
for ${s(s+1) \over 2} - 1 < m \leq {(s+1)(s+2) \over 2} - 1$.
\\ \indent
Returning to eq. (\ref{OverEll}), with the $l_i$ set to $2, 3, \ldots, s+1$,
	\begin{equation*}
		c(\Theta) > 2^{m-1}/(s+1)!.			
	\end{equation*}
Since $m > \frac{s(s+1)}{2} - 1$, $s < \frac{\sqrt{8m + 9} - 1}{2} < \sqrt{2m}$ (the latter for $m>1$). 
Since $(s+1)! > 2 s^s$, we find the bound on the number of cycles for $m$ nodes in loops (for $m\geq1$) is
	\begin{equation}
		c(m) > 
		2^{m - 2 - \sqrt{2m} \log_2\sqrt{2m}} = O(2^m).			
		\label{AttractorNumber}
	\end{equation}
\noindent
{\sf\textbf{\textcolor{purple}{Discussion}}} \\
We proved that the number of cycles scales as $2^{m}$.
This is considerably faster than $2^{0.5m}$ derived in \cite{Flyvbjerg88} via a lengthier calculation,
and $2^{0.47m}$ derived in \cite{Drossel05}.
\\ \indent
The lower bound on the number of attractors gives an upper bound on the mean attractor length.
Writing  $d(\Theta) = \nu_1 \ci A_1 + \nu_2 \ci A_2 + \ldots$, where there are $\nu_i$ cycles of length $A_i$,
the mean attractor length is
	\begin{equation}
		\overline{A} = \sum_{i} \nu_i A_i \Big/  \sum_{i} \nu_i = \sum_{i} \nu_i A_i \Big/ c(m).
		\label{Conservation}
	\end{equation}
Since all $2^m$ states of the loop nodes belong to cycles, $\sum_{i} \nu_i A_i = 2^m$.
Inserting this and eq. (\ref{AttractorNumber}) into eq. (\ref{Conservation}), the mean attractor length satisfies
\begin{eqnarray*}
  \overline{A}(m) < \sqrt{2m}^{\sqrt{2m}} = 2^{\sqrt{2m} \log_2\!\sqrt{2m}},
\end{eqnarray*}
much slower than $2^{0.5m}$ and $2^{0.53m}$ derived in \cite{Flyvbjerg88} and \cite{Drossel05}.
\\ \indent
Now let's re-express eq. (\ref{AttractorNumber}) in terms of the number of nodes in the network $N$, whereby $m$ becomes a random variable.
In the large $N$ limit, the mean number of loops of length $l$ is $\exp\left(-l^2/(2N)\right)/l$.
Summing over this, the mean number of nodes in loops $\overline{m}$ is asymptotically $\sqrt{{\pi \over 2} N}$.
Since this is convex, by Jensen's inequality
	\begin{equation*}
		c(N) > 2^{1.25 \sqrt{N}},
	\end{equation*}
compared to the best known $2^{0.63 \sqrt{N}}$ and $2^{0.59 \sqrt{N}}$ \cite{Flyvbjerg88, Drossel05}.
\\ \indent
Key to our approach is our formalism for expressing the dynamics of multiple loops as a product of the dynamics of individual loops.
It opens the door to further insight by turning a problem about dynamics into a problem about combinatorics.
If, as is widely believed, all critical Kauffman models behave in a similar way \cite{Drossel05}, 
our insights into the critical model with connectivity one apply to other critical models as well.

T. Fink thanks Andriy Fedosyeyev for useful advice.

\end{document}